\documentclass{aa}  

\usepackage{graphicx}
\usepackage{natbib}
\usepackage{txfonts}
\usepackage{lipsum}
        
\usepackage{lscape}            
\usepackage{placeins}

\usepackage{xcolor}

\begin{document}

   \title{On-sky demonstration of a vector Zernike wavefront sensor in a cascaded adaptive optics system}

   \author{M. Motte\inst{1,2}\fnmsep\thanks{Corresponding author: mathieu.motte@lam.fr}
   \and V. Chambouleyron\inst{2}
   \and R. Fétick\inst{1,2}
   \and F. Oyarzun\inst{2} 
   \and M.A. Alagao\inst{2} 
   \and A. Striffling\inst{3} 
   \and E. Vinerskas \inst{2,1}
   \and J.-F. Sauvage\inst{1,2}
   \and C.T. Héritier\inst{1,2}
   \and E. Muslimov\inst{4} 
   \and M. Cissé\inst{5}
   \and A. Rahim\inst{6}  
   \and J. Kent Wallace \inst{7}
   \and T. Wenger \inst{7}
 \and B. Neichel\inst{2} 
 \and T. Fusco\inst{1,2}
      }

   \institute{DOTA, ONERA, 13330, Salon-de-Provence, France \
             \and  Aix Marseille University, CNRS, CNES, LAM, Marseille, France\
             \and Aix Marseille Univ, CNRS, Pytheas, OHP, Observatoire de Haute-Provence, France \
            \and Department of Physics, University of Oxford, Keble Rd, OX14 3RH Oxford \
            \and W. M. Keck Observatory, 65-1120 Mamalahoa High \
            \and Université Côte d’Azur, Observatoire de la Côte d’Azur, CNRS, Laboratoire Lagrange, Nice, France
            \and Jet Propulsion Laboratory, California Institute of Technology, Pasadena, CA 91109, USA
       }

   \date{}

  \abstract
   {Future high-contrast instruments aim to directly image and characterise Earth-like exoplanets. Achieving the required contrast requires exquisite image quality, and in particular a high-performance correction of the optical aberrations. Hence the future instrument will be using adaptive optics systems, operating at increasingly high loop frequencies in order to reduce temporal errors. However, increasing the loop frequency decreases the number of photons available per wavefront-sensor frame, motivating the use of highly sensitive wavefront sensors. Among the available concepts, the Zernike wavefront sensor (ZWFS) approaches the theoretical sensitivity limit of Fourier-filtering wavefront sensors. To extend the limited dynamic range of the classical ZWFS, the vector Zernike wavefront sensor (v-ZWFS) has been proposed as an alternative solution to mitigate this limitation. Although ZWFSs have already been employed for non-common-path aberration compensation and segment phasing, they have not yet been demonstrated in an on-sky closed-loop adaptive optics system.}
   {We aim to demonstrate the operation as well as the performance of a Zernike wavefront sensor in an on-sky adaptive optics loop and to evaluate its potential as a second-stage wavefront sensor in a cascaded adaptive optics architecture.} 
   {We implement a second-stage adaptive optics system, named OZIRIIS, on the PAPYRUS adaptive optics platform installed on the 1.52~m telescope of the Observatoire de Haute-Provence. OZIRIIS combines a vector Zernike wavefront sensor and a 97-actuator deformable mirror operating at 400~Hz downstream of the first-stage pyramid-based adaptive optics system. We use one single ZWFS signal for real-time control. The full v-ZWFS is only used a posteriori to analyse the telemetry. We develop a complete calibration strategy based on a numerical model of the sensor to generate synthetic reference signals and interaction matrices for both laboratory and on-sky operation. The performance of the system is assessed through point spread function (PSF) image analysis and wavefront sensor telemetry and is then compared with simulations using a numerical twin of the bench.}
   {We present the on-sky demonstration of a vector Zernike wavefront sensor operating within an adaptive optics system, together with on-sky closed-loop results obtained using one of the ZWFS signals for real-time control. The second-stage correction provides an improvement in the measured Strehl ratio of up to 16 percentage points during the observing night presented here. The analysis of the telemetry using the full v-ZWFS to reconstruct residuals reveals the optical-gain effects affecting the ZWFS at low Strehl ratio. This demonstrates that the v-ZWFS can improve the dynamic range and performance of the ZWFS in regimes with larger aberrations. We show that an accurate model of the sensor enables the generation of synthetic reference signals and interaction matrices suitable for on-sky operation. We further show that the observed behaviour is consistent with numerical simulations and that the impact of optical gain can be quantified directly from the telemetry.}
    {}


   \maketitle
    \nolinenumbers
\section{Introduction} \label{sec:intro} 

\begin{figure*}[!t]

  \centering
  \includegraphics[width=\textwidth]{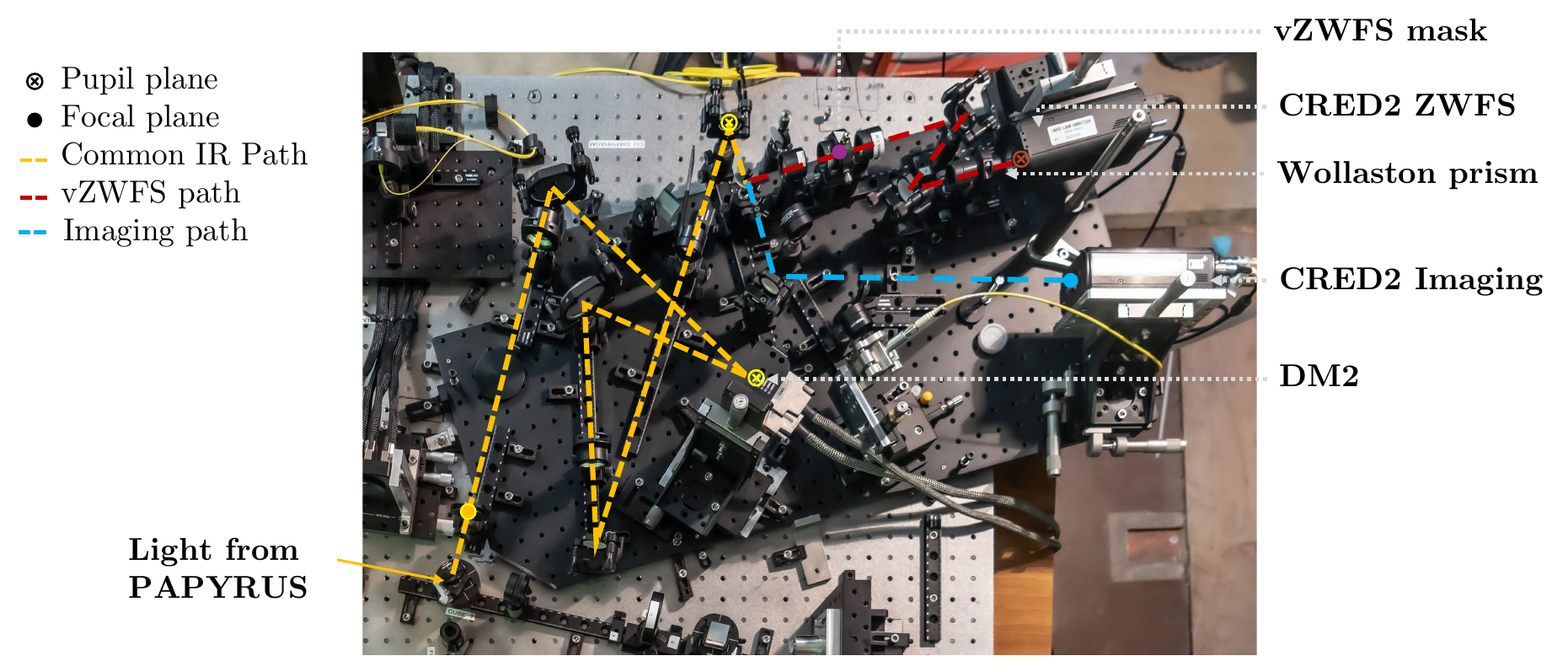}
  \caption{Top view of the IR path of PAPYRUS with OZIRIIS.}
  \label{fig:OZIRIIS}
 \end{figure*}

Since the first direct images of exoplanetary systems, the imaging and spectroscopic characterisation of Earth-like exoplanets has become one of the primary goals of exoplanetary science. Such observations could provide valuable insights into the formation of planetary systems \citep{Haffert19}, the thermodynamics and chemistry of planetary atmospheres \citep{Bugatti2025}, and, ultimately, the detection of potential biomarkers \citep{Schwieterman2018}. Any residual or uncorrected wavefront error that degrades the electromagnetic field produces a light signature in the image, known as a speckle, which can lead to false detections. Achieving the required contrast and angular resolution for ground-based observatories, hence, demands eXtreme Adaptive Optics correction of all sources of phase aberrations, including atmospheric turbulence, segment piston errors, structural vibrations, dome seeing, chromatic effects, etc.

Experience gained with first-generation high-contrast imaging instruments such as SPHERE \citep{Beuzit2019} and GPI \citep{Macintosh2014} has shown that temporal error is one of the dominant error sources in such systems, particularly at low spatial frequencies, thereby limiting performance in the immediate vicinity of the star where the targets of interest are located \citep{Cantalloube2019SPHEREContrast, Cantalloube2020, Bailey2016}. Reducing this error requires operating the adaptive optics (AO) loop at higher frequencies. However, increasing the loop frequency reduces the number of photons detected per frame and consequently the signal-to-noise ratio (SNR) of the wavefront sensor (WFS) measurements. Employing highly sensitive WFSs may provide a solution to this limitation, but their reduced dynamic range makes it challenging to directly correct atmospheric turbulence. A potential solution is therefore to deploy such sensors within a dual-stage AO architecture to correct the low-order residual aberrations left uncorrected by a single AO stage \citep{Urra2022, Kasper2021}. In this configuration, the first stage provides robust turbulence correction, while the second stage, owing to its higher sensitivity, can run faster and improve the correction of the system. This dual-stage concept is being implemented in instruments under development, such as SAXO+ \citep{Goulas2024} and GMagAO-X \citep{Haffert2022, Kautz23}.

In this context, the Zernike wavefront sensor (ZWFS) appears to be a promising solution. It builds on the phase-contrast principle introduced by \citet{Zernike1934}. In the regime of small aberrations, the core of the focal-plane field, once isolated, behaves as a flat wavefront. It can therefore serve as a reference beam that interferes with the surrounding field. This is achieved by focusing the stellar light onto a small phase-shifting mask that applies a phase shift to the core of the electromagnetic field without significantly affecting the surrounding field. After propagation to the pupil plane, the resulting intensity pattern encodes the sine of the phase. The first advantage of the ZWFS is its sensitivity, which approaches the theoretical limit of Fourier-filtering wavefront sensors \citep{Chambouleyron2021}. Furthermore, phase discontinuities such as differential piston and the low wind effect (LWE) are readily measured by the ZWFS, unlike with commonly used WFSs such as the Shack--Hartmann and modulated pyramid wavefront sensors \citep{NDiaye2016,Cisse2022}.

For these reasons, the ZWFS has already been deployed in several astronomical instruments. On the SPHERE instrument, the ZELDA mask is able to measure the non-common-path aberrations (NCPA) and characterise the low wind effect \citep{Sauvage2015LowWindEffect, NDiaye2016}. A ZWFS has also been operated on Keck II to measure segment piston errors \citep{vanKooten2022}. In-lab, the GHOST bench experiment demonstrated the potential of the ZWFS for correcting residual aberrations from a first AO stage \citep{Ndiaye2024}. Building on these results, the ZWFS is now being considered for several high-contrast instruments under development, including the Asgard--Baldr interferometer on the VLTI \citep{Taras24,Courtney24}, RISTRETTO on the VLT \citep{Shinde2024}, and, in the longer term, PCS on the ELT \citep{Kasper2021}.

The phase-shifted ZWFS (PSZWFS) uses two ZWFSs with different phase-shift depths. The diversity between the two sensor signals can double the dynamic range of a single ZWFS, at the expense of a loss in sensitivity to read-out noise by a factor of $\sqrt{2}$ \citep{Cisse2022}. The vector ZWFS is a practical implementation of this concept, in which a single metasurface mask is used to generate the two ZWFS signals. The metasurface introduces two different phase shifts to the two orthogonal polarisations, with phase-shift depths chosen at the fabrication stage of the metasurface. Once the two polarisations are separated, two ZWFS pupil images corresponding to two different phase-shift depths are obtained \citep{cisse2023, Chambouleyron2024, wallace2023}. Recently, the Keck implementation was upgraded to a vector ZWFS (v-ZWFS) to increase its dynamic range and successfully correct differential piston errors between segments \citep{Salama2024}.

Despite these developments, to our knowledge, no Zernike wavefront sensor has yet been demonstrated on sky as part of a closed-loop AO system. Since its first light in 2022, one of the objectives of the PAPYRUS adaptive optics platform, installed on the 1.52~m telescope at the Observatoire de Haute-Provence, has been to test new AO concepts for high-contrast imaging instruments (Alagao et al., in prep.). We used this platform to perform the first on-sky demonstration of a Zernike wavefront sensor operating within a fast AO loop. In this paper, we introduce OZIRIIS, the second-stage AO system of PAPYRUS, which is based on a vector Zernike wavefront sensor. We describe the design of the instrument, its laboratory characterisation, and its expected performance. We then present the on-sky results obtained with a single ZWFS signal used for real-time control. Exploiting the full v-ZWFS signal requires a non-linear reconstruction scheme, which was not implemented in real time in the present work; the full v-ZWFS information is therefore used a posteriori to assess the performance of the system. Lastly, we discuss the lessons learnt for future implementations.

\begin{figure}[!t]

  \centering
  \includegraphics[width=\columnwidth]{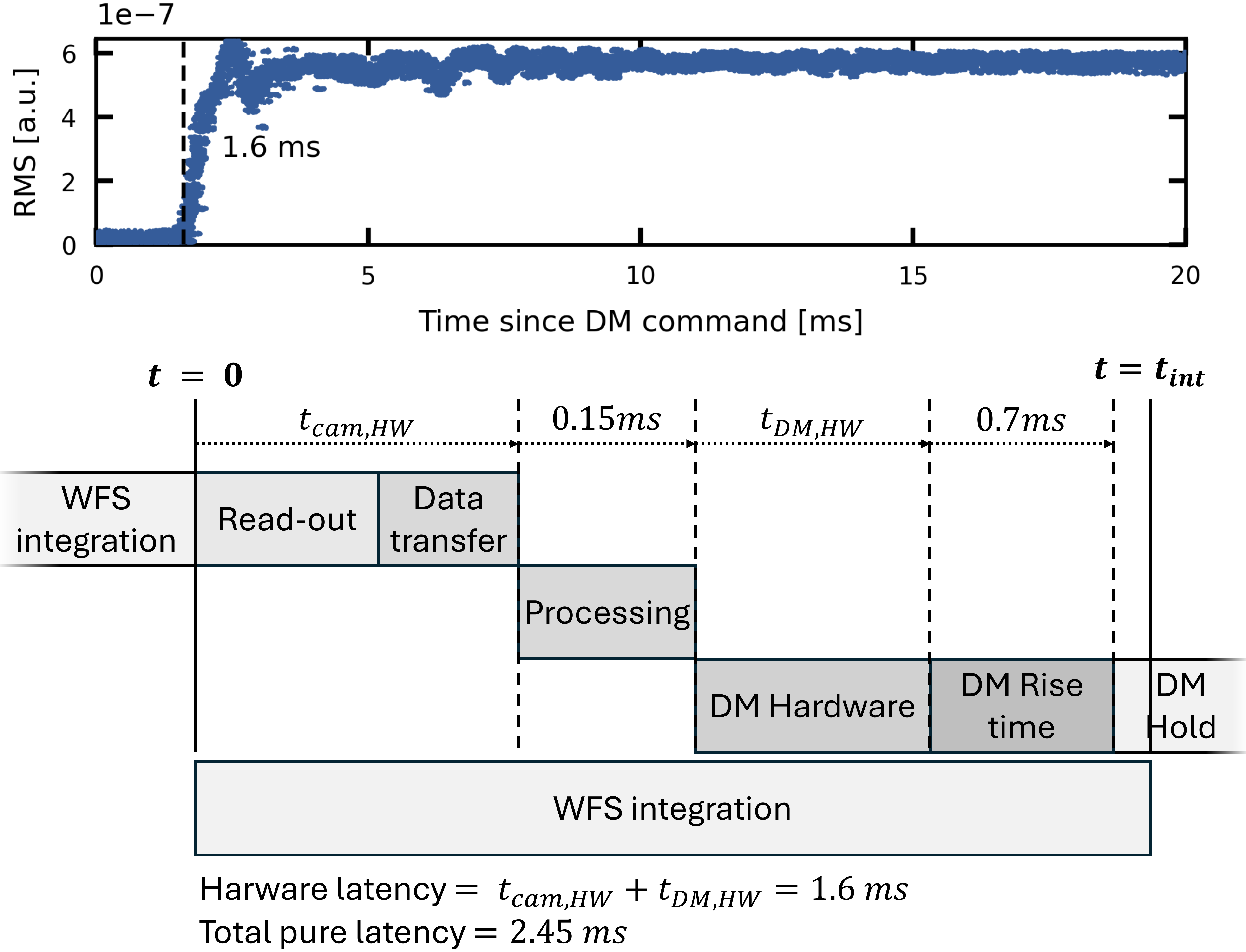}
  \caption{Top: temporal evolution of the RMS of the WFS images after a command is written in the DM shared memory. The measured hardware latency is 1.6~ms. Bottom: chronogram of the OZIRIIS adaptive optics system.}
  \label{fig:chronogram}
 
\end{figure} 
\section{The PAPYRUS cascaded AO system}

\subsection{PAPYRUS first AO stage}

The design and performance of PAPYRUS are described in detail in \citet{Alagao2026PAPYRUS}. PAPYRUS is an AO bench installed on the Coudé focus of the 1.52~m telescope at the Observatoire de Haute Provence in southern France. The median seeing measured at T152 is about $2.5''$.

After the common path, the wavelengths below 950\,nm are reflected toward the first AO stage. The first stage consists of a modulated four-sided glass pyramid (mPWFS) used to control a 241 actuators running nominally at 500~Hz. It achieves an average J-band SR of $20.9\%$ \citep{Alagao2026PAPYRUS}.

The infrared branch is mounted on a 400~mm $\times$ 800~mm breadboard located in one corner of the visible bench, as shown in Fig.~\ref{fig:OZIRIIS}. All wavelengths higher than 950~nm are transmitted towards this branch. It hosts an infrared imaging camera, a fibre-injection module feeding the VIPA spectrograph \citet{carlotti2022sky}, and the second-stage AO system OZIRIIS, described in detail in the following section.
A calibration unit is located upstream of PAPYRUS and provides internal illumination through a 635~nm laser source for the visible channel and a $1550\,\mathrm{nm}$ laser source for the infrared channel.

\subsection{OZIRIIS: PAPYRUS second AO stage} \label{section:OZIRIIS}

OZIRIIS (On-sky Zernike WFS InfraRed II\textsuperscript{nd} Stage) is the second AO stage of PAPYRUS and is installed on the infrared branch. Its optical design is shown in Fig.~\ref{fig:OZIRIIS}. After the first-stage dichroic, infrared light enters the common optical path of the second stage with an F\# of 14.8. The beam is collimated and the pupil is imaged onto the second-stage DM2. A second dichroic then separates the infrared light between the imaging and wavefront-sensing branches. The cut-off wavelength is about $1.55\,\mu\mathrm{m}$. Shorter wavelengths are directed towards the CRED2 infrared imaging camera, while longer wavelengths are sent to the wavefront-sensing branch or to the VIPA spectrograph injection module.
\begin{table}
\caption{Configuration parameters of the dual-stage PAPYRUS system.}
\label{tab:ao_configuration}
\centering
\renewcommand{\arraystretch}{1.15}
\begin{tabular}{lcc}
\hline\hline
Parameter & First stage & Second stage \\
\hline
Sensor & mPWFS $5\lambda/D$ & ZWFS \\
Pixels in pupil diameter & 72 & 84\\
DM & $15 \times 15$ & $9 \times 9$ \\
Controlled modes & 195 & 35\\
Frame rate (Hz) & 200 & 400 \\
Pure frame delay & 0.8 & 1.0 \\
Integrator gain & 0.7 & 0.3 \\
Leak & 0.995 & 0.980 \\
Spectral band (nm) & $\leq950$& $1550$ to $1700$\\
\hline
\end{tabular}
\end{table}
\begin{figure*}[!htbp]

  \centering
  \sidecaption
  \includegraphics[width=0.68\textwidth]{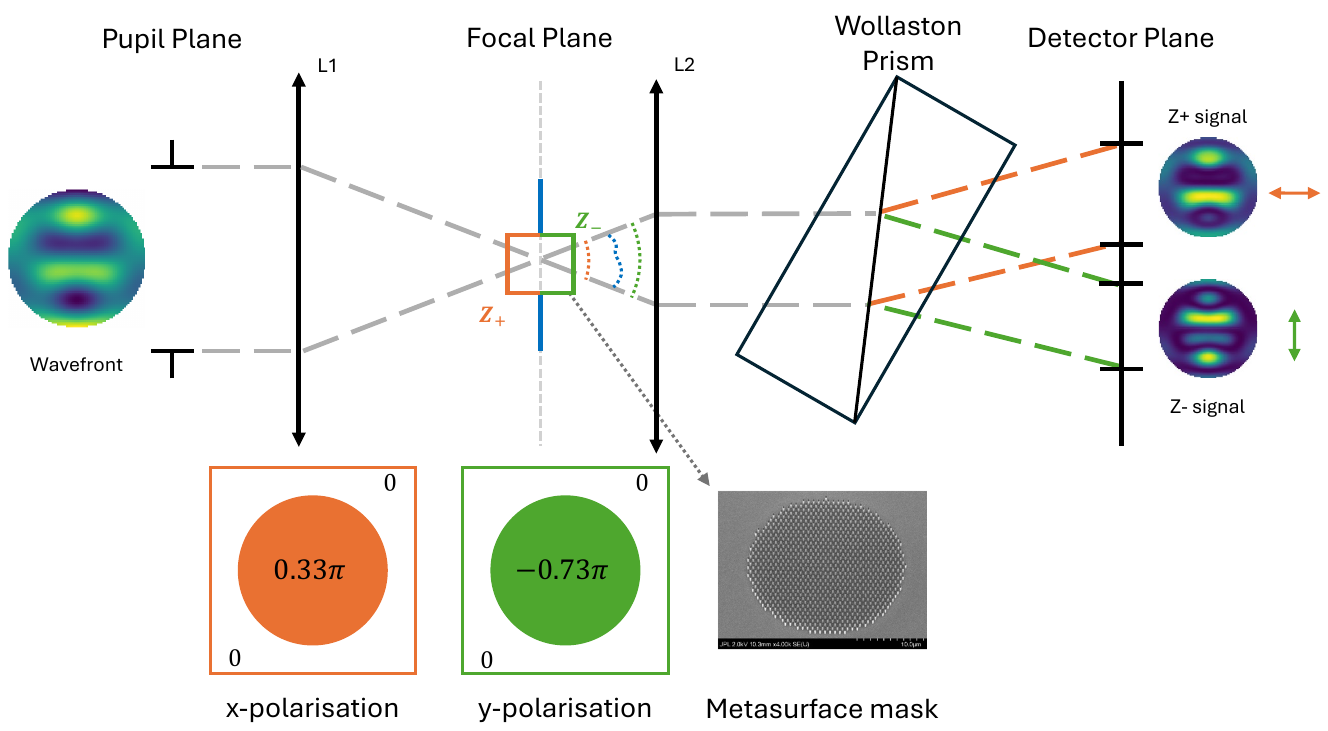}
  \caption{Schematic representation of the v-ZWFS concept. The metasurface mask from \citet{wallace2023} applies two different phase shifts to two orthogonal polarisations. A Wollaston prism then separates the two polarisations, allowing the two ZWFS pupil images to be recorded. The bottom-right inset shows an SEM image of the metasurface used in this work, \citet{wallace2023}.}
  \label{fig:vZWFS_scheme}
 
\end{figure*}
The second AO stage is made of a metasurface v-ZWFS \citep{Wenger25}, it is described in detail in section \ref{subsection:v-ZWFS}. The DM2 was initially conceived for NCPA compensation and is not designed to operate above 400 Hz. To emphasise the performance gain provided by the second stage, and to reproduce a dual-stage AO architecture similar to those envisaged for future VLT and ELT instruments, the first-stage loop frequency is reduced from 500 Hz to 200 Hz, while OZIRIIS is operated at 400 Hz. Both AO loops are controlled by the DAO real-time controller \citep{barr_2025_17264152}. The two AO stages configuration is detailed in Tab. \ref{tab:ao_configuration}. 
A detailed understanding of the temporal behaviour of the system is essential for interpreting its on-sky performance. We therefore measured the latency of the OZIRIIS control loop in order to interpret its temporal rejection and to build the second-stage error budget used in the simulations. Excluding the detector integration time and the DM zero-hold, the pure loop latency is 2.45~ms, corresponding to approximately 0.98 frames at 400~Hz. To estimate the total system latency, we first measure the hardware latency, defined as the elapsed time between the instant a command is written to the DM shared memory and the reception in a shared memory of the first WFS frame containing the response to this command \citep{Guyon2018}. This measurement is performed by applying a periodic command to the DM at a given frequency while acquiring WFS images asynchronously with respect to the DM command stream. By computing the RMS difference between each ZWFS frame and a reference frame acquired at (t=0), and then unfolding the resulting time series, a super-resolved temporal response of the DM as recorded by the camera can be reconstructed (Fig.~\ref{fig:chronogram} - top). The hardware latency is subsequently determined as the time interval between (t=0) and the onset of the DM response. Using this approach, the hardware latency is estimated to be 1.6~ms. The DM rise time, specified by ALPAO, is approximately 0.7 ms. In the present analysis, this response time is modelled as an equivalent pure delay. The RTC processing time is obtained from software-generated timestamps and is found to be approximately 0.15 ms. A detailed chronogram summarising all the latency contributions described above is presented in Fig.~\ref{fig:chronogram}. 
In the on-sky configuration, the first-stage AO system corrects up to 195 Karhunen--Loève (KL) modes, while OZIRIIS corrects up to 35 modes. Larger number of modes led to reduced closed-loop stability. The number of corrected modes was therefore limited by closed-loop stability rather than by the number of available DM degrees of freedom.

\subsection{Vector Zernike wavefront sensor}\label{subsection:v-ZWFS}

Due to its limited dynamic range, the ZWFS can only be operated in the weak-aberration regime. Furthermore, when the amplitude of the aberrations becomes too large, the core of the focal-plane electromagnetic field no longer behaves as a planar reference wavefront and is not suitable to be used as a reference for the interferometric measurement. To extend the linearity range while preserving the high sensitivity of the ZWFS, one can employ a variant known as the phase-shifted ZWFS. The principle consists in combining the signals from two ZWFSs with different phase-shift depths. \citet{Cisse2022} and \citet{Chambouleyron2024} implemented this concept using a vector Zernike wavefront sensor (v-ZWFS), first introduced by \citet{Doelman19} for simultaneous phase and amplitude reconstruction, and demonstrated that the two complementary signals can be exploited to extend the dynamic range of the classical ZWFS. 

Our vector Zernike mask is manufactured using metasurfaces \citep{Wenger25}. The principle of the v-ZWFS is detailed in Fig.~\ref{fig:vZWFS_scheme}. A Wollaston prism is used to introduce an angular separation between the two orthogonal polarisations, allowing two complementary ZWFS signals to be recorded. In practice, an additional relay lens is inserted after the Zernike mask to adjust the magnification imposed by the camera sampling, the available mask size, and the commercial Wollaston prism.

The Wollaston prism also introduces an ellipticity in the two pupil images and an optical path difference between the two polarisation states, resulting in two distinct pupil planes. In the on-sky configuration considered here, the loop is closed using only one of the two ZWFS signals. The camera is therefore focused on the selected pupil image, which depends on the mask parameters.

Two aspects of the Wollaston-prism alignment are important. First, in agreement with the analysis of \citet{Simon86}, the Wollaston prism must be intentionally tilted with respect to the optical axis, as this configuration significantly reduces the ellipticity of the pupil images. Second, accurate beam collimation is required to preserve pupil circularity, as residual beam divergence produces elliptical pupil images. In the working spectral band, the chromaticity of the Wollaston prism does not significantly affect the quality of the on-sky pupil images. The v-ZWFS pupil images are sampled with a resolution of $90\times90$ pixels in the calibration configuration and $84\times84$ pixels during on-sky operation. The latter diameter is limited by M1 mirror or T152, while the former is limited by the oversized DM of 1st stage. 

The main challenge of the v-ZWFS is that a simple interaction-matrix inversion is not sufficient to recover the phase, as it remains limited to the linear regime of the two individual ZWFS signals. Exploiting the full dynamic range therefore requires non-linear reconstructors such as those proposed by \citet{Haffert2024} and \citet{Chambouleyron2024}. These approaches are considered too computationally expensive for real-time operation in a high-speed AO system. After this study, \citet{Oyarzun2026CNNvZWFS} demonstrated, using OZIRIIS, the feasibility of real-time phase reconstruction using a v-ZWFS with a convolutional neural networks. 

In the present work, we focus on the demonstration of on-sky closed-loop operation of a single ZWFS within an AO loop. Although the instrument provides two complementary v-ZWFS signals, only one of them is used for real-time control of the second-stage AO loop. The commands sent to the DM are reconstructed by applying the pseudo-inverse of the interaction matrix to the selected ZWFS signal. The signal $s$ used for the reconstruction is defined as: 
\begin{equation}
s = \frac{I_{\mathrm{ZWFS}}}{\sum I_{\mathrm{ZWFS}}}-\frac{I_{\mathrm{ref}}}{\sum I_{\mathrm{ref}}},
\end{equation}
where $I_{\mathrm{ZWFS}}$ is the intensity measured by the CRED2 camera over the valid pixels of the selected ZWFS image, and $I_{\mathrm{ref}}$ is the reference intensity, the setpoint towards which the sensor signal must converge.
The on-sky closed-loop results obtained with this real-time single-ZWFS control strategy are presented in section ~\ref{section:Results}. In the following, we describe the non-linear full v-ZWFS reconstruction used a posteriori to analyse the residual phase and assess the performance of the system.

As shown by \citet{Chambouleyron2024}, gradient-descent approach and iterative arctangent reconstructors achieve similar reconstruction accuracy. We therefore employ the iterative arctangent reconstructor throughout this work. The equations used for the arctangent reconstructor have already been described in \citet{Haffert2024} and \citet{Chambouleyron2024}. Let us consider two ZWFSs with phase-shift depths $\theta_1$ and $\theta_2$, and detector intensity distributions $I_1$ and $I_2$, respectively. Here, $I_k$ denotes the intensity measured in the $k$-th ZWFS pupil image on the detector, with $k \in \{1,2\}$. Let $\phi$ be the phase of the electromagnetic field to be measured. We denote by $I_{p,k}$ the pupil intensity distribution on the detector, corresponding to the pupil image that would be obtained without the focal mask, and by $I_{b,k}$ and $\beta_k$ the intensity distribution and phase of the reference field on the detector plane, respectively. This reference field corresponds to the contribution produced by the part of the focal-plane field that passes through the phase-mask dimple of the $k$-th ZWFS. In practice, $I_{p,k}$ can be obtained either by removing the mask or by moving the PSF outside the mask. $I_{b,k}$ can be computed by propagating the field transmitted through a focal-plane binary amplitude mask that has the same diameter as the ZWFS dimple, transmits only the region covered by the dimple, and blocks the surrounding field. The arctangent reconstructor can then be written as:

\begin{equation}
\begin{pmatrix}
s_1 \\
s_2
\end{pmatrix}
=
\begin{pmatrix}
-\sin\left(\beta_1+\theta_1/2\right) & \cos\left(\beta_1+\theta_1/2\right) \\
-\sin\left(\beta_2+\theta_2/2\right) & \cos\left(\beta_2+\theta_2/2\right)
\end{pmatrix}
\begin{pmatrix}
\cos\phi \\
\sin\phi
\end{pmatrix},
\end{equation}
where
\begin{equation}
s_k =
\frac{
I_k - I_{p,k} - 4 I_{b,k} \sin^2\left(\theta_k/2\right)
}{
4\sqrt{I_{b,k}I_{p,k}}\sin\left(\theta_k/2\right)
}.
\end{equation}
By inverting this $2\times2$ system, one obtains $\cos\phi$ and $\sin\phi$. The phase is then reconstructed as:
\begin{equation}
\hat{\phi} =
\operatorname{atan2}
\left(
\sin\phi,
\cos\phi
\right).
\end{equation}

The phase map $\beta_k$ of the field passing through the mask is unknown, as it depends on the phase $\phi$ that we aim to measure. Therefore, in the iterative arctangent reconstructor, we first assume $\beta_k=0$ and reconstruct a phase around this initial reference. This reconstructed phase is then used to compute an updated estimate of $\beta_k$, from which a new phase estimate is obtained. By iterating this procedure approximately ten times, the wavefront phase can be accurately retrieved \citep{Chambouleyron2024}.

In summary, two reconstruction methods are used in this work, but in different contexts: (i) a linear reconstruction based on the interaction matrix of a single ZWFS signal is used for real-time control of the bench and for closing the adaptive optics loop; and (ii) the iterative arctangent reconstructor, using the full v-ZWFS signal, is applied a posteriori to analyse the second-stage telemetry after the observations, as it is not suitable for real-time reconstruction in the present implementation.

\subsection{Characterisation of the v-ZWFS mask} \label{subsection:charact}

In this section, we aim to obtain an accurate characterisation of the v-ZWFS in order to build a reliable model of the sensor. This is required for two reasons: (i) the iterative arctangent reconstruction relies on precise knowledge of the phase shifts introduced by the masks and of the corresponding reference signals; and (ii) the model is used to generate the reference signals themselves. Let $\psi$ be the input electromagnetic field, $M$ the ZWFS mask, $\theta$ the phase-shift depth of the mask, and $H$ the binary top-hat function describing the mask. The field in the detector plane, $\psi_{\mathrm{det}}$, can then be written as
\begin{equation}
\psi_{\mathrm{det}}
=
\mathrm{FT}^{-1}\left(\hat{\psi} \, M\right)
=
\psi
-
\left(1 - e^{i\theta}\right)
\mathrm{FT}^{-1}\left(\hat{\psi} \, H\right).
\end{equation}

This equation shows that the Fourier transform only needs to be computed in the focal-plane region covered by the mask. This can be done using a matrix Fourier transform, as already described for Roddier and Lyot coronagraphs by \citet{Soummer07}. This modelling principle for the ZWFS has been implemented in the OOPAO Python adaptive optics simulator \citep{Heritier2023}, and the v-ZWFS model is obtained simply by combining the two ZWFSs.

It follows that the two main parameters of a ZWFS are the mask diameter relative to the PSF, which defines $H$, and the phase-shift depth $\theta$. The accuracy of this model therefore depends primarily on the calibration of these two parameters. In the following, we describe how both the effective mask diameter and the phase shift depths are estimated.

The diameter of the v-ZWFS mask is measured using a microscope and estimated to be 23$\mu$m. Traditionally, the diameter of the ZWFS mask in the focal plane is chosen to be $1.06\lambda/D$, however, \citet{Chambouleyron2021} show that a diameter of $2\lambda/D$ significantly improved the sensitivity of the ZWFS. Consequently, the optical design was adjusted to obtain an effective mask diameter as close as possible to $2\lambda/D$. Given the constraints imposed by the available off-the-shelf optical components and the fixed physical size of the metasurface mask, the final value is $1.96\lambda/D$ in the on-sky configuration. The pupil seen in calibration is a circular unobstructed aperture that is approximately 6\% larger than the on-sky pupil. In addition, the calibration wavelength differs slightly from the central wavelength used for wavefront sensing on-sky. As a result, the effective mask diameter becomes $2.14\lambda/D$ when using the internal calibration source. While the mask diameter can be measured directly, the phase depths of the two masks must be estimated from the sensor response on calibration source.

\subsubsection{Depth and diameter of the mask}

The phase depths of the two v-ZWFS masks are characterised by comparing measured ZWFS images with synthetic images generated from the v-ZWFS model. The procedure is summarised below.

\begin{enumerate}
\item For each ZWFS signal $k\in\{1,2\}$, select a range of trial phase-shift depths $\theta_k$.
\item For each trial value $\theta_k$, reconstruct a phase estimate from the measured ZWFS image, using a single ZWFS non-linear arcsine reconstructor \citep{Steeves2020}, used here only as an intermediate calibration step.
\item Propagate this phase estimate through the v-ZWFS model to generate a synthetic detector image $I_{\mathrm{synth},k}(p;\theta_k)$.
\item Compute the least-squares error
\begin{equation}
\chi_k^2(\theta_k)
=
\sum_p
\left[
I_{\mathrm{meas},k}(p)
-
I_{\mathrm{synth},k}(p;\theta_k)
\right]^2 ,
\end{equation}
where $p$ denotes the pixel index.
\item Select the phase-shift depth that minimises this error,
\begin{equation}
\theta_k^\star
=
\operatorname*{argmin}
\chi_k^2(\theta_k).
\end{equation}
\end{enumerate}

The minimisation is applied independently to the two ZWFS signals. The phase-shift depth corresponding to the minimum least-squares error is retained for each signal.

\begin{figure}[!htbp]

  \centering
  \includegraphics[width=\columnwidth]{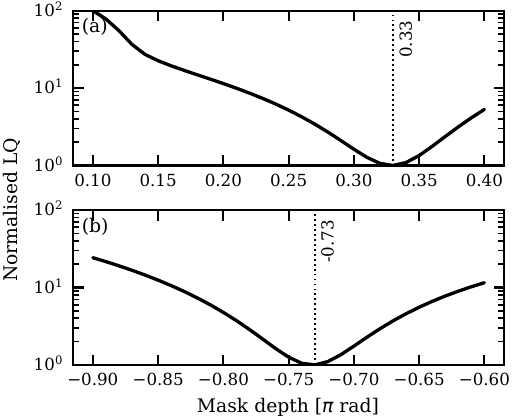}
  \caption{Normalised least-squares error of the difference between simulated images and observed images for the two ZWFS pupils. The minima correspond to phase-shift depth values of $0.33\pi$ and $-0.73\pi$.}
  \label{fig:characterisation}
 
\end{figure}

Using this approach, the phase-shift depths were estimated to be $0.33\pi$ and $-0.73\pi$ for the two polarisations, as shown in Fig.~\ref{fig:characterisation}. One of the two measured phase shifts is therefore relatively close to the nominal $\pi/2$ phase shift, while the phase difference between the two ZWFS signals is approximately $1.06\pi$. The optimal v-ZWFS configuration in terms of sensitivity is known to correspond to phase shifts of $-0.5\pi$ and $0.5\pi$ \citep{Chambouleyron2024}, yielding a phase difference of $\pi$. These phase shifts were the target values for the metasurface mask. The measured deviation from these target values is attributed to fabrication tolerances in the metasurface manufacturing process. The measured values are therefore expected to reduce the sensitivity and slightly reduce the dynamic range of the sensor, although the resulting v-ZWFS still provides a larger dynamic range than a single ZWFS. Furthermore, as one mask is close to $\pi$, it ejects more light outside the pupil and is expected to be less linear. We therefore decided to focus the camera on the pupil image of the $0.33\pi$ ZWFS.

\begin{figure}[t]
  \centering
  \includegraphics[width=\columnwidth]{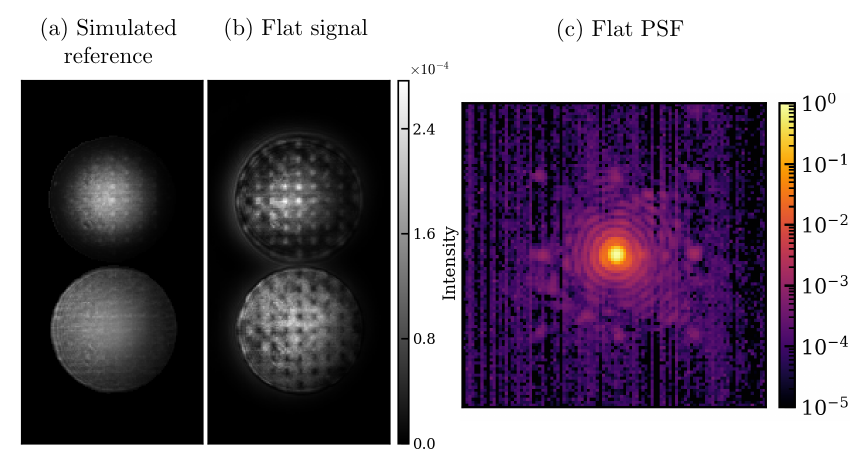}
  \caption{(a) Simulated v-ZWFS reference signal. (b) Experimental v-ZWFS signal obtained after flattening the second-stage wavefront using the simulated reference signal. In panels (a) and (b), the top pupil image corresponds to the $-0.73\pi$ ZWFS, while the bottom pupil image corresponds to the $0.33\pi$ ZWFS used for real-time AO control. (c) PSF measured on the bench with the flattened DM shape shown in panel (b). The measured Strehl ratio is estimated at 71\% at 1550~nm.}
  \label{fig:Best_flat_signal}
\end{figure}

In order to converge to a flat wavefront, the ZWFS reference signal must be known. This can be obtained experimentally either by flattening the wavefront using an external wavefront sensor or by using PSF images acquired with the calibration source. However, this procedure is time-consuming. More importantly, the on-sky reference is expected to differ from the one obtained in calibration and cannot be accessed through this experimental method. This motivates the use of a synthetic reference, which can be generated from the calibrated v-ZWFS model. To do so, a large tilt is applied to the second-stage DM, moving the PSF outside the v-ZWFS mask. In this configuration, the CRED2 camera records the pupil intensity distribution. This pupil image is then propagated through the v-ZWFS model to generate the corresponding reference signal. Fig.~\ref{fig:Best_flat_signal} presents the second-stage flat obtained using this simulated reference. The resulting PSF measured on the science camera corresponds to a Strehl ratio of approximately $71\%$ at 1550~nm. After NCPA compensation on the bench, a Strehl ratio of $79\%$ was obtained on the PAPYRUS imaging camera (Alagao et al., in prep.). This indicates that the static error budget is not entirely due to NCPA: the residual common-path aberration is estimated to be approximately $120\,\mathrm{nm}$ RMS, including a DM print-through contribution of about $30\,\mathrm{nm}$ RMS, while the NCPA contribution is estimated to be approximately $80\,\mathrm{nm}$ RMS. The print-through can be clearly observed as a waffle pattern at twice the maximum frequency on the bench flat signal on Fig.~\ref{fig:Best_flat_signal}b and as high order spatial frequencies visible on the PSF in Fig.~\ref{fig:Best_flat_signal}c. This waffle pattern also appears in the simulated reference signal as one of the pupils is not exactly at focus and therefore produces an intensity change by Talbot effect. We also tested an alternative reference signal in which the residual phase reconstructed by the v-ZWFS after bench flattening was included in the input phase of the v-ZWFS model. In this context, this residual phase denotes the phase map measured by the v-ZWFS around the flattened bench state, rather than a separate estimate of a specific aberration term. Including this residual phase in the reference did not improve the closed-loop stability in bench tests and occasionally degraded it. A possible explanation is that it introduces a bias in the reference signal, so that the loop is closed around a non-zero reconstructed phase rather than around a flat wavefront, thereby reducing the usable linear range. Since this additional term did not improve the loop behaviour, we used a synthetic reference signal generated from a flat wavefront for closed-loop operation.

The v-ZWFS model can therefore provide a suitable reference for the observations and can be used in an on-sky context. This approach is particularly useful because it is both adaptable and fast. In our case, this is especially important because the structure of the reference wavefront differs slightly between the calibration and on-sky configurations. This difference is mainly due to the presence of the central obstruction in the on-sky pupil and to the change in pupil size. The resulting on-sky reference signal is shown in Fig.~\ref{fig:sky_ZWFS}. The effect of the central obstruction on the reference signal is clearly visible, especially for the $-0.73\pi$ ZWFS.
\subsubsection{Dynamic range of the sensor}
\begin{figure}[!t]

  \centering
  \includegraphics[width=\columnwidth]{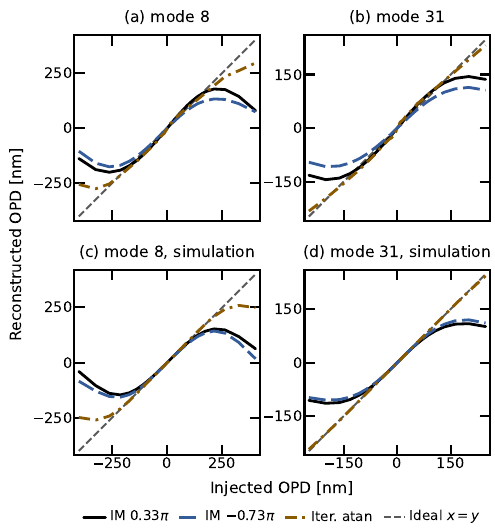}
  \caption{Top: experimental linearity curves of the ZWFSs and the v-ZWFS with an iterative arctangent reconstruction, Bottom: simulated linearity curves of the ZWFSs and the v-ZWFS with an iterative arctangent reconstruction. $IM$ $-0.73\pi$ and $IM$ $0.33\pi$ correspond respectively to IM of the $-0.73\pi$ and $0.33\pi$ signals. These measurements were obtained using the calibration source, so for a wavelength of $1550\,\mathrm{nm}$ and a mask size of 2.14$\lambda/D$.}
  \label{fig:ZWFS_linearity}
\end{figure}
To estimate the performance of the v-ZWFS in an on-sky configuration, we measure the linearity curves of the two individual ZWFS masks and of the combined v-ZWFS. Using the calibrated v-ZWFS model, we also compared the experimental results with numerical simulations. To obtain the linearity curves, each mode is successively applied to the DM with increasing amplitude. The resulting aberrations are then reconstructed using the interaction matrices of the two individual ZWFSs. For the v-ZWFS, the phase is reconstructed using the iterative arctangent algorithm. The reconstructed modal amplitudes are finally compared to the injected amplitudes. The resulting linearity curves are presented in Fig.~\ref{fig:ZWFS_linearity}.
For low-order modes, the individual ZWFSs provide accurate measurements up to approximately 200~nm RMS, while for higher-order modes the linear range decreases to about 150~nm RMS. The v-ZWFS is able to reconstruct aberrations up to approximately 300~nm RMS, corresponding to about twice the range achieved by an individual ZWFS for high-order modes.

Interestingly, the measured dynamic ranges of the $0.33\pi$ and $-0.73\pi$ ZWFSs are more similar in the model than on the bench. This discrepancy is explained by the detector position. The CRED2 camera is focused on the pupil plane corresponding to the $0.33\pi$ mask, and thus the pupil associated with the $-0.73\pi$ mask is slightly defocused on the detector. As a consequence, the phase aberrations measured by the $-0.73\pi$ mask are only partially converted into intensity fluctuations on the bench, reducing its effective phase-to-intensity conversion gain. This relative defocus is not taken into account in the model. It is important to note that these measurements were performed using the calibration source, at a wavelength of $1550\,\mathrm{nm}$ and with a mask size of $2.14\lambda/D$. In the on-sky configuration, the effective mask size is $1.96\lambda/D$. Since the dimple diameter affects the modal sensitivity of the ZWFS and can degrade non-linear PSZWFS reconstruction when it becomes too large \citep{Chambouleyron2021,Chambouleyron2024}, the calibration linearity curves should not be interpreted as an exact measurement of the on-sky linearity range.
Overall, these results confirm that the v-ZWFS provides a substantially larger dynamic range than a classical ZWFS and remains in good agreement with the theoretical model. The final objective is to take benefit from this extended dynamic range in real time using a non-linear reconstructor. Those results are presented in \citet{Oyarzun2026CNNvZWFS}. 

\section{On-sky results}\label{section:Results}
\subsection{AO calibration strategy}

The calibration procedure mainly consists in obtaining a reference signal, described in section \ref{subsection:charact}, and in having an accurate interaction matrix. PAPYRUS provides two internal calibration sources. In principle, the infrared calibration source operating at 1550~nm could be used to measure the interaction matrix directly. However, this approach presents several limitations in the case of OZIRIIS. As discussed previously, the calibration pupil differs from the on-sky pupil both in size and in shape. Consequently, the PSF and the corresponding ZWFS response are not identical in the two configurations. Furthermore, the calibration path introduces ghost images at the v-ZWFS wavelength, while vibrations affecting the PAPYRUS bench degrade the quality of the measured interaction matrix. These limitations, combined with the accurate knowledge of the v-ZWFS parameters, motivate the use of a synthetic interaction matrix computed in simulation. 

\begin{figure}[!ht]

  \centering
  \includegraphics[width=\columnwidth]{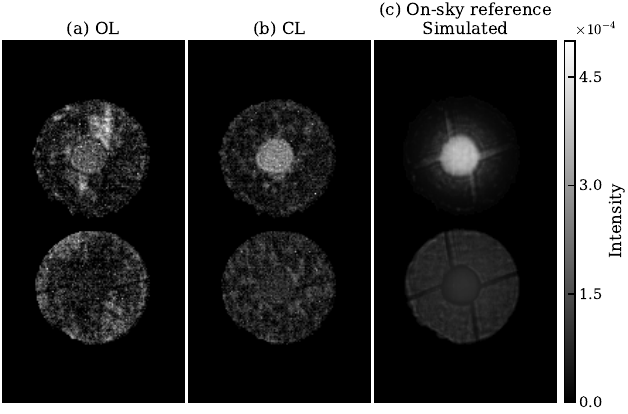}
  \caption{On-sky CRED2 frames acquired with the second-stage loop open (left) and closed (centre). The right panel shows the reference signal used for real-time closed-loop control. In each panel, the upper and lower pupil images correspond to the v-ZWFS channels with phase shifts of $-0.73\pi$ and $0.33\pi$, respectively. Only the $0.33\pi$ channel was used for real-time AO control.}
  \label{fig:sky_ZWFS}
\end{figure}

The quality of a synthetic interaction matrix strongly depends on the accuracy of the DM model. Both the influence functions and the DM misregistrations must therefore be known precisely. In particular, errors in the estimated translations, rotations, or magnification between the DM actuators and the WFS pupil can significantly affect the reconstruction accuracy and loop stability. The influence functions are directly given by ALPAO. The misregistrations are estimated using the SPRINT algorithm developed by \citet{Heritier2021}. To do so, an experimental interaction matrix is first acquired on the bench and compared with a synthetic interaction matrix. The algorithm then fit the DM degrees of freedom required to reproduce the experimental measurements.

Once the misregistrations are identified, a new KL modal basis is computed using the calibrated DM model. This basis must account for the actual on-sky pupil geometry, including the central obstruction and the pupil offset relative to the calibration configuration. We find that the ZWFS is particularly sensitive to edge effects and can rapidly exhibit instabilities near the pupil boundaries and the central obstruction. Computing the KL basis, and therefore the mode-to-command matrix, using the correct pupil geometry mitigates these effects and avoids actuator slaving.

With accurate models of both the v-ZWFS and the DM, the complete system can then be simulated using OOPAO \citep{Heritier2023} to compute the synthetic interaction matrix. An additional advantage of synthetic interaction matrices is that they can be generated directly during observations. Before each observing sequence, a large tilt is applied to the DM to move the PSF outside the v-ZWFS mask and measure the pupil intensity distribution. This pupil image is propagated through the v-ZWFS model to generate both the reference signal (Fig.~\ref{fig:sky_ZWFS}c) and the corresponding synthetic interaction matrix. It is important to note that both of them are computed assuming monochromatic light at the central wavelength of the on-sky wavefront-sensing band.
\subsection{AO performance}

\begin{figure}[t!]

  \centering
  \includegraphics[width=\columnwidth]{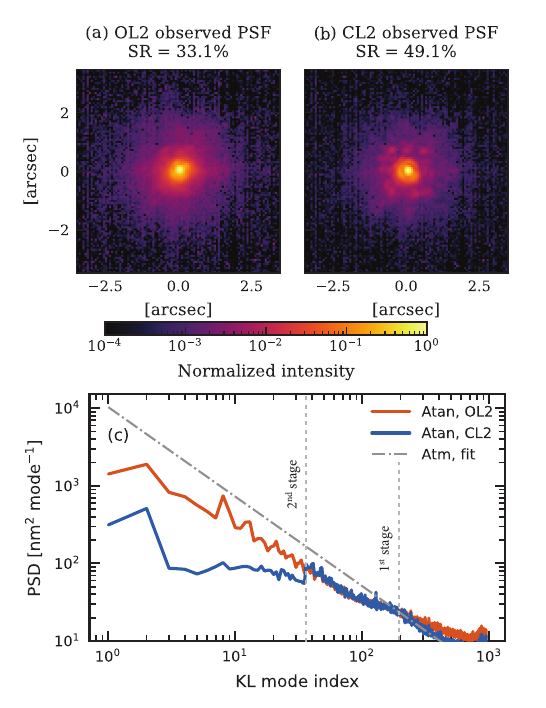}
  \caption{Top: open-loop and closed-loop PSFs observed on 16 April 2026 at 23:05 and 23:03 UTC, respectively. Bottom: modal PSDs of the corresponding second-stage telemetry. The PSF-based Strehl ratios were obtained using a polychromatic PSF model and are reported at an effective wavelength of $\lambda_{\mathrm{eff}} = 1350\,\mathrm{nm}$.}
  \label{fig:psfs}
 \end{figure}

\begin{figure*}[t!]

  \centering
  \sidecaption
  \includegraphics[width=0.7\textwidth]{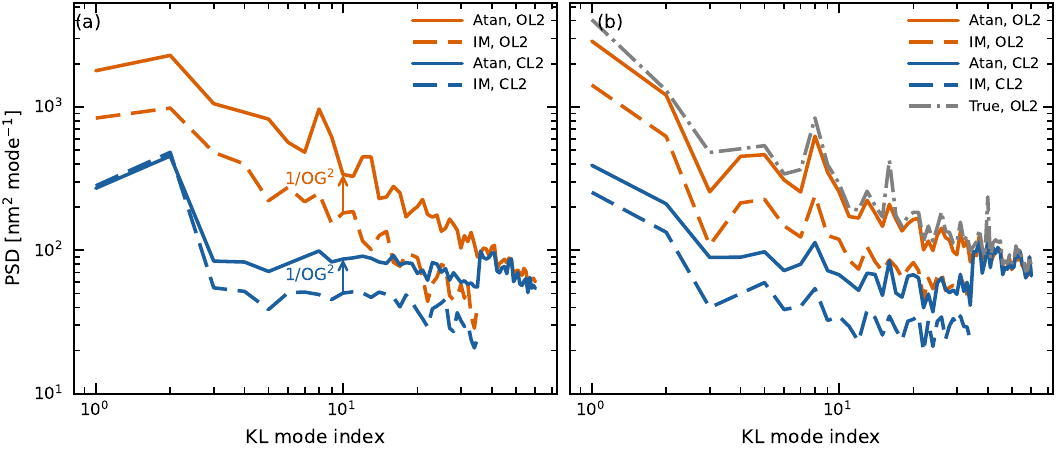}
  \caption{Spatial PSDs: (a) On-sky PSDs obtained by reducing telemetry with linear reconstructor (IM) and iterative atan reconstructor. (b) PSDs obtained in simulation under the same AO conditions.}
  \label{fig:modal_psds}
 \end{figure*}

OZIRIIS was operated on several nights between February and April 2026. In this section, we present a representative observing sequence obtained on 16 April 2026 at 23:00 UTC on HD98262 ($H=0.18$) at a zenith angle of $Z = 20^\circ$. The seeing measured under full open-loop conditions is approximately $2.4''$ at 550~nm. As detailed in Sect.~\ref{section:OZIRIIS} and Table~\ref{tab:ao_configuration}, the first-stage loop frequency is set to 200~Hz and the second-stage loop frequency to 400~Hz. The first-stage loop gain is set to 0.7, above its nominal value to compensate for optical-gain effects, with a leak factor of 0.995 and a pyramid modulation radius of $5\lambda/D$. Only the $0.33\pi$ ZWFS signal is used for real-time control of the second-stage AO loop. The full v-ZWFS information is recorded and used a posteriori to assess the performance of the system. We note that fine tuning the gains of the two stages to optimise the overall system performance is beyond the scope of this paper. 

The data are first acquired with OZIRIIS operating in closed loop at 23:03 UTC, followed by open-loop measurements at 23:05 UTC. Throughout the sequence, the first-stage AO loop remains closed. For each configuration, we record the complete second-stage telemetry, including the DM2 shapes and commands, the v-ZWFS images and reconstructed commands, the infrared PSF images, and the associated timestamps. The PSF images are acquired at 400~Hz, yielding a total of 10000 frames for each dataset.

Figure~\ref{fig:sky_ZWFS} presents a representative v-ZWFS CRED2 frame acquired with the second-stage loop open, a representative frame acquired in closed loop, and the reference signal used for wavefront reconstruction. The closed-loop correction is performed using an interaction matrix built from the bottom ZWFS signal only. A clear reduction of the low-order aberrations is observed in closed loop. Using the full v-ZWFS telemetry images, the iterative arctangent reconstructor is then applied a posteriori to reconstruct the residual phase and generate the corresponding long-exposure PSF. The v-ZWFS reconstruction yields Strehl ratios of $45\%$ in open loop and $61\%$ in closed loop at $1600\,\mathrm{nm}$. When the estimated differential NCPA contribution is included, these values correspond to Strehl ratios of approximately $29\%$ and $44\%$, respectively, at the effective wavelength used for the Strehl-ratio analysis of the observed PSFs, $\lambda_{\mathrm{eff}} = 1350\,\mathrm{nm}$. The Strehl ratios derived from the observed PSFs were computed using the polychromatic PSF model implemented in MAOPPY \citep{fetick2019}. The resulting values are consistent with those estimated independently from the v-ZWFS telemetry.

The PSFs shown in Fig.~\ref{fig:psfs} illustrate the correction of low-order aberrations provided by the ZWFS and the associated reduction of the temporal residual error. From these PSF measurements, the second-stage correction provides a Strehl-ratio gain of approximately $16$ percentage points. This behaviour is further confirmed by the modal power spectral densities (PSDs) reconstructed a posteriori using the iterative arctangent reconstructor. The PSDs clearly reveal the correction regions of both AO stages and demonstrate efficient correction of the low-order modes. 

Using the commands reconstructed by the real-time controller at each iteration, we derive the residual phase estimated by the linear reconstructor and project it onto the modal basis. This allows a direct comparison between the real-time linear reconstruction and the a posteriori iterative arctangent reconstruction. The resulting modal PSDs are presented in Fig.~\ref{fig:modal_psds}a. We observe that using the interaction matrix on the bottom ZWFS signal systematically underestimates the phase variance. \citet{Ndiaye2013} showed that the optical gains of the ZWFS can be approximated by the square root of the Strehl ratio (SR). Since the PSDs are expressed in variance units, the ratio between the PSDs reconstructed with the linear and arctangent reconstructors is expected to scale as the square of the optical gains, and therefore approximately as the SR. We measure average ratios of $0.4$ in open loop and $0.55$ in closed loop between the PSDs reconstructed with the linear and arctangent reconstructors. PSDs ratios are consistent with the observed SRs, suggesting that the discrepancy between the two reconstructors is primarily due to optical gain effects. It is important to note that we assume the iterative reconstructor to be less affected by optical-gain effects. By construction, this reconstructor is non-linear and iteratively recomputes the reference field around which the phase measurement is performed. However, this does not imply that it is fully insensitive to optical gain. At large residual phase amplitudes, the iterative arctangent reconstructor may no longer estimate the reference field accurately, and optical-gain-like effects may still affect the reconstruction. 

Figure~\ref{fig:modal_psds}b presents the corresponding numerical simulation obtained under similar seeing conditions. A sequence of $5000$ atmospheric turbulence screens is first generated at $400$~Hz and injected into the PAPYRUS numerical twin implemented in OOPAO \citep{Heritier2023}, operating at $200$~Hz with the same control parameters as those used during the observations. The residual phase delivered by the first-stage simulation is then injected into the OZIRIIS numerical model, which includes both the DM and the v-ZWFS models. The second-stage loop is operated using the same parameters as during the on-sky observations. At each iteration, the commands reconstructed by the linear reconstructor and the corresponding v-ZWFS signals are recorded. Following the same procedure as for the on-sky telemetry, the residual phase is reconstructed a posteriori using the iterative arctangent reconstructor and compared with both the real-time linear reconstruction and the theoretical residual phase. We observe that the behaviour measured on-sky is well reproduced by the simulation. The iterative arctangent reconstructor recovers a phase PSD that closely matches the theoretical residual phase, whereas the interaction-matrix reconstruction systematically underestimates the phase variance. This discrepancy is consistent with the expected optical-gain effects and therefore supports the interpretation of the on-sky measurements and indicates that the iterative arctangent reconstruction is not significantly affected by the same linear optical-gain attenuation in this regime.

\begin{figure*}[ht!]

  \centering
  \sidecaption
  \includegraphics[width=0.7\textwidth]{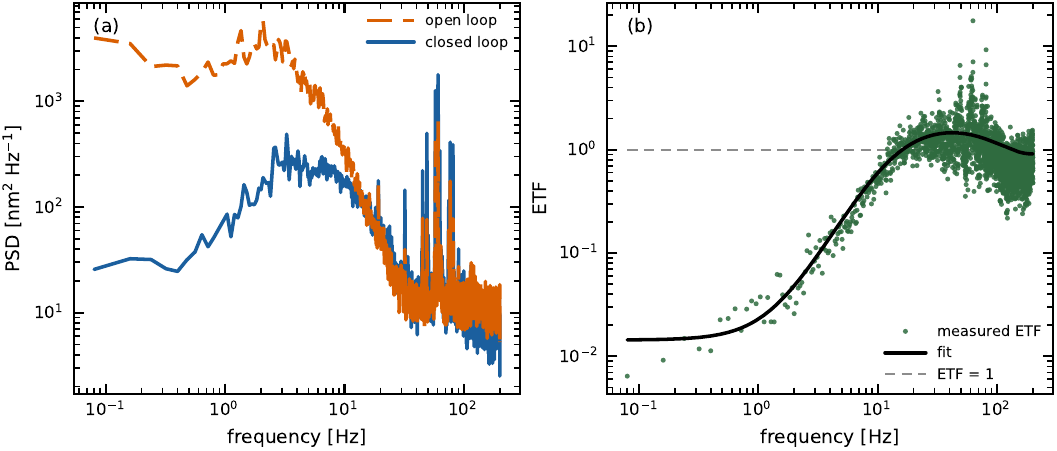}
  \caption{Temporal PSDs and ETFs from AO telemetry, observations of the April 16, 2026 closed loop at 23:03 UTC and open loop at 23:05 UTC.}
  \label{fig:etf}
 \end{figure*}

The close agreement between the on-sky and simulated results indicates that the main physical mechanisms governing the system behaviour are adequately captured by the numerical model. It further suggests that the bench calibration procedures and the associated error budget are well understood. This agreement provides confidence in the predictive capability of the simulation framework and supports its use for assessing the expected performance of the system under observing conditions that were not directly explored during the on-sky campaign.

\begin{figure}[ht!]

  \centering
  \includegraphics[width=\columnwidth]{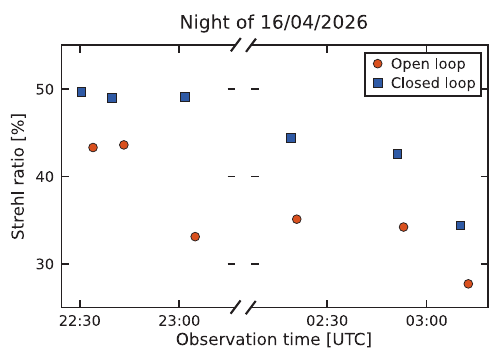}
  \caption{Strehl ratio vs observation time UTC in open and closed loop of the second stage. Before midnight the target is HD98262 ($H =0.18$, $Z = 20^\circ$), and in the late night Arcturus ($H =-2.8$, $Z = 40^\circ$).}
  \label{fig:SR}
 \end{figure}

To characterise the temporal behaviour of the system, we compute the temporal PSD of each actuator and derive the error transfer function (ETF) of the AO control loop from the reconstructed commands and their associated timestamps. The sum of the temporal PSDs of the actuators is shown in Fig.~\ref{fig:etf}. The control loop efficiently rejects frequencies below 15~Hz. Above 30~Hz, the PSD becomes noise dominated, although vibration peaks remain visible in both open-loop and closed-loop data around 50~Hz. The corresponding ETF is displayed in the right panel of Fig.~\ref{fig:etf}. Since the system latency is known, the ETF can be fitted with a discrete-time control model. At a loop frequency of 400~Hz, the pure frame delay corresponds to approximately 0.97 frames. The fit yields an effective loop gain of 0.15, approximately half of the gain applied in the controller. This discrepancy can be explained by two effects. First, the optical gain of the ZWFS is expected to scale as the square root of the Strehl ratio \citep{Ndiaye2013}. For the measured Strehl ratio of 61\%, this corresponds to a factor of $\sqrt{0.61}\simeq0.78$. Second, a posteriori analysis revealed a mismatch between the synthetic interaction matrix and the measured interaction matrix of the bench, resulting in an additional gain factor of approximately 0.7. Accounting for both effects, the expected effective gain is $0.3\times\sqrt{0.61}\times0.7\simeq0.164$, in good agreement with the value obtained from the ETF fit. This agreement indicates that the two dominant contributors to the loop-gain discrepancy are correctly identified and quantified. It further supports our interpretation of the on-sky measurements and provides confidence that the dominant terms of the second-stage AO error budget are correctly captured.

After presenting one observing sequence in detail, we now consider all sequences for which both open-loop and closed-loop PSF measurements were acquired during this observing night. Figure~\ref{fig:SR} shows that closing the second-stage loop consistently improves the measured Strehl ratio, with gains ranging from 5 to 16\% over the night. This confirms the measurement made with the ZWFS and demonstrates that the second AO stage improves the performance delivered by the first stage.

\section{Discussion and conclusion}

In this work, we have presented the first on-sky demonstration of a Zernike wavefront sensor operating in closed loop within a fast adaptive optics system. We introduced OZIRIIS, the second-stage AO system of PAPYRUS, and described its design, calibration strategy, and on-sky operation. Although OZIRIIS is based on a vector Zernike wavefront sensor, the real-time control presented in this paper relies on a single ZWFS signal, while the full v-ZWFS information is exploited a posteriori for performance analysis.

This demonstration relied on a detailed characterisation of the vector Zernike wavefront sensor. The resulting calibrated model was shown to provide accurate reference signals and synthetic interaction matrices for both laboratory and on-sky conditions. The measurements also confirmed the significantly larger dynamic range offered by the v-ZWFS compared to a classical ZWFS. In the present work, this extended dynamic range was used only for a posteriori analysis, while a single ZWFS signal was employed for real-time control. The use of the full v-ZWFS signal in a real-time closed loop with a CNN-based reconstructor is addressed in a separate work by \citet{Oyarzun2026CNNvZWFS}.

The on-sky observations demonstrated stable operation of the second-stage AO loop under the challenging seeing conditions of the Observatoire de Haute-Provence. Throughout the observing night, we systematically observed an improvement in Strehl ratio with the second-stage loop closed. The comparison between the linear and iterative arctangent reconstructions revealed the impact of the ZWFS optical gain. The good agreement between on-sky measurements and numerical simulations confirmed the validity of the numerical model and showed that the dominant error terms of both PAPYRUS and OZIRIIS are well understood. The temporal analysis further showed that the discrepancy between the commanded and effective loop gains can be explained by the combined effects of optical gain and interaction-matrix uncertainties.

It is important to note that the control of the two-stage AO system was not fully optimised. In this work, we used a classical cascaded AO architecture, and the optimisation of the loop gains of both stages could be further improved. The implementation of a disentangled control architecture could also improve the overall performance of the system \citep{galland2023}.

Several aspects of the model and calibration strategy could still be improved. In particular, we identified a mismatch between the synthetic and on-bench interaction matrices, which contributes to the reduction of the effective loop gain. This mismatch may arise from residual modelling uncertainties in the DM response, pupil registration, or optical propagation. In addition, while the ZWFS operates over a finite spectral band, the current model is monochromatic. The impact of chromaticity was not investigated in this work, but could be relevant for broadband operation with a ZWFS.

The use of a synthetic interaction matrix was primarily motivated by the fact that the internal calibration unit does not accurately reproduce the on-sky configuration. Additional motivations include the presence of ghost images and vibrations in the calibration path, which degrade the quality of experimentally measured interaction matrices. If an instrument provides a calibration unit with the correct pupil geometry and an appropriate spectral band, a measured interaction matrix may be preferable, since modelling uncertainties can reduce the efficiency of the AO loop. However, even in that case, synthetic reference signals may remain valuable, as they can provide a significant time saving compared to experimental wavefront-flattening techniques.

The optical implementation of a v-ZWFS also raises specific instrumental considerations. In the present implementation, the detector was aligned with the pupil image of the ZWFS signal used for on-sky closed-loop operation. This choice was appropriate for real-time control with a single ZWFS signal, but it may not be optimal for exploiting the full v-ZWFS signal with a non-linear reconstructor. Future implementations should therefore investigate the optimal detector positioning in the presence of the Wollaston prism, and in particular whether both pupil images should be brought into the same conjugation plane to fully benefit from the v-ZWFS information. During on-sky operation, we did not observe any polarisation effects induced by the telescope or the instrument that affected the system performance. However, before implementing a v-ZWFS in a future instrument, polarisation effects should be carefully considered during the design phase.

The ZWFS is among the most sensitive wavefront sensors and can measure phase discontinuities such as segment piston, petalling modes, and the low-wind effect \citep{NDiaye2016,Cisse2022,Salama2024}. This study demonstrates that a ZWFS can correct dynamic aberrations in an on-sky configuration and that adding a ZWFS-based second AO stage downstream of an existing AO system can lead to a measurable performance improvement. This result highlights the potential of Zernike wavefront sensing for cascaded AO architectures dedicated to high-contrast imaging. Although our system was limited to 400~Hz by the second-stage deformable mirror, the high sensitivity of the ZWFS makes it suitable for operation in faster AO systems. Together with the possibility of extending its dynamic range through vector implementations, this makes the ZWFS a promising candidate for future extreme-AO instruments such as PCS on the ELT \citep{Kasper2021}.

\begin{acknowledgements}
This work benefited from the support the French National Research Agency (ANR) with the Programme Investissement Avenir F-CELT (ANR-21-ESRE-0008), PEPR ORIGINS (XAO-WFS ANR-22-EXOR-00062 and COMPACT SPECTROGRAPH ANR-22-EXOR-0006), the ANR-DGA-AID ASTRID program (ANR-25-ASTR-0015), the CNRS-IRP GALILEAO program, the Action Spécifique Haute Résolution Angulaire (ASHRA) of CNRS/INSU co-funded by CNES, the french government under the France 2030 investment plan (cassiopée project) and the Initiative d’Excellence d’Aix-Marseille Université A*MIDEX, program number AMX-22-RE-AB-151.
\end{acknowledgements}

\bibliographystyle{aa} 
\bibliography{report}

\end{document}